\documentclass[12pt]{article}
\title{Improved randomized selection}
\author{Krzysztof C. Kiwiel\thanks{Systems Research Institute,
        Newelska 6, 01--447 Warsaw, Poland
        ({\tt kiwiel@ibspan.waw.pl})}}
\date{February 2, 2004}

\newcommand{\eqref}[1]{{\normalfont\normalcolor(\ref{#1})}}
\makeatletter
\def\proof{%
   \def\a##1{\begin{trivlist}\item[]{\bf\ignorespaces{##1}.}%
    \enspace\ignorespaces}%
   \def\b[##1]{\a{Proof\ \ignorespaces{##1}}}%
   \@ifnextchar[{\b}{\a{Proof}}}
\def\endproof{\end{trivlist}}
\def\qed{\relax\protect\ifmmode\ifinner\else\quad\fi\fi
    \hbox{\vbox{\hrule height.4pt\hbox{\vbox{\hrule height.4pt
    \hbox{\vrule width.4pt\vphantom{\normalsize A}\kern.5em
    \vrule width.4pt}\hrule height.4pt}}}}}
\newtoks\@stequation
\def\subequations{\refstepcounter{equation}%
\edef\@savedequation{\the\c@equation}%
\@stequation=\expandafter{\theequation}%   %only want \theequation
\edef\@savedtheequation{\the\@stequation}% %expanded once
\edef\oldtheequation{\theequation}%
\setcounter{equation}{0}%
\def\theequation{\oldtheequation\alph{equation}}}%
\def\endsubequations{%
\setcounter{equation}{\@savedequation}%
\@stequation=\expandafter{\@savedtheequation}%
\edef\theequation{\the\@stequation}\global\@ignoretrue}
\def\@begintheorem#1#2{\trivlist
    \item[\hskip \labelsep{\bfseries #1\ #2.}]\itshape}
\def\@opargbegintheorem#1#2#3{\trivlist
    \item[\hskip \labelsep{\bfseries #1\ #2\ (#3).}]\itshape}
\@addtoreset{equation}{section}% Makes \section reset `equation' counter.
\def\theequation{\thesection.\arabic{equation}}
\@addtoreset{figure}{section}

\@addtoreset{table}{section}

\let\@@eqnsel=\relax
\def\@tempa{%
    \stepcounter{equation}%
    \def\@currentlabel{\p@equation\theequation}%
    \global\@eqnswtrue\m@th
    \global\@eqcnt\z@
    \tabskip\mathindent
    \let\\=\@eqncr
    \setlength\abovedisplayskip{\topsep}%
    \ifvmode
      \addtolength\abovedisplayskip{\partopsep}%
    \fi
    \addtolength\abovedisplayskip{\parskip}%
    \setlength\belowdisplayskip{\abovedisplayskip}%
    \setlength\belowdisplayshortskip{\abovedisplayskip}%
    \setlength\abovedisplayshortskip{\abovedisplayskip}%
    $$\everycr{}\halign to\linewidth% $$
    \bgroup
      \hskip\@centering
      $\displaystyle\tabskip\z@skip{##}$\@eqnsel&%
      \global\@eqcnt\@ne \hskip \tw@\arraycolsep \hfil${##}$\hfil&%
      \global\@eqcnt\tw@ \hskip \tw@\arraycolsep
        $\displaystyle{##}$\hfil \tabskip\@centering&%
      \global\@eqcnt\thr@@
        \hb@xt@\z@\bgroup\hss##\egroup\tabskip\z@skip\cr}%
\def\@tempb{%
   \stepcounter{equation}%
   \def\@currentlabel{\p@equation\theequation}%
   \global\@eqnswtrue
   \m@th
   \global\@eqcnt\z@
   \tabskip\@centering
   \let\\\@eqncr
   $$\everycr{}\halign to\displaywidth\bgroup
       \hskip\@centering$\displaystyle\tabskip\z@skip{##}$\@eqnsel
      &\global\@eqcnt\@ne\hskip \tw@\arraycolsep \hfil${##}$\hfil
      &\global\@eqcnt\tw@ \hskip \tw@\arraycolsep
         $\displaystyle{##}$\hfil\tabskip\@centering
      &\global\@eqcnt\thr@@ \hb@xt@\z@\bgroup\hss##\egroup
         \tabskip\z@skip
      \cr
}
\ifx\eqnarray\@tempa%     If the fleqn document-class option is in effect
    \def\eqnarray{%
    \stepcounter{equation}%
    \def\@currentlabel{\p@equation\theequation}%
    \global\@eqnswtrue\m@th
    \global\@eqcnt\z@
    \tabskip\mathindent
    \let\\=\@eqncr
    \setlength\abovedisplayskip{\topsep}%
    \ifvmode
      \addtolength\abovedisplayskip{\partopsep}%
    \fi
    \addtolength\abovedisplayskip{\parskip}%
    \setlength\belowdisplayskip{\abovedisplayskip}%
    \setlength\belowdisplayshortskip{\abovedisplayskip}%
    \setlength\abovedisplayshortskip{\abovedisplayskip}%
    $$\everycr{}\halign to\linewidth% $$
    \bgroup
      \hskip\@centering
      $\displaystyle\tabskip\z@skip{##}$\@eqnsel&%
      \global\@eqcnt\@ne
      \@@eqnsel%            \@@eqnsel has replaced \hskip \tw@\arraycolsep!!!
      \hfil${{}##{}}$\hfil&%              as in fixup.sty but textstyle!!!
      \global\@eqcnt\tw@
      \@@eqnsel%           \@@eqnsel has replaced \hskip \tw@\arraycolsep!!!
        $\displaystyle{##}$\hfil \tabskip\@centering&%
      \global\@eqcnt\thr@@
        \hb@xt@\z@\bgroup\hss##\egroup\tabskip\z@skip\cr}%
\else\ifx\eqnarray\@tempb%       Else try the default eqnarray environment.
   \def\eqnarray{%
   \stepcounter{equation}%
   \def\@currentlabel{\p@equation\theequation}%
   \global\@eqnswtrue
   \m@th
   \global\@eqcnt\z@
   \tabskip\@centering
   \let\\\@eqncr
   $$\everycr{}\halign to\displaywidth\bgroup
       \hskip\@centering$\displaystyle\tabskip\z@skip{##}$\@eqnsel
      &\global\@eqcnt\@ne
      \@@eqnsel%           \@@eqnsel has replaced \hskip \tw@\arraycolsep!!!
      \hfil${{}##{}}$\hfil%              as in fixup.sty but textstyle!!!
      &\global\@eqcnt\tw@
      \@@eqnsel%           \@@eqnsel has replaced \hskip \tw@\arraycolsep!!!
         $\displaystyle{##}$\hfil\tabskip\@centering
      &\global\@eqcnt\thr@@ \hb@xt@\z@\bgroup\hss##\egroup
         \tabskip\z@skip
      \cr}
\else 
\fi\fi
\def\@tempa{}			% Free up TeX's memory
\def\@tempb{}
\@ifundefined{chapter}{%
  \renewenvironment{thebibliography}[1]
     {\section*{\refname
        \@mkboth{\MakeUppercase\refname}{\MakeUppercase\refname}}%
      \list{\@biblabel{\@arabic\c@enumiv}}%
           {\settowidth\labelwidth{\@biblabel{#1}}%
            \leftmargin\labelwidth
            \advance\leftmargin\labelsep
            \itemsep \z@                 % Suppresses vertical separation.
            \@openbib@code
            \usecounter{enumiv}%
            \let\p@enumiv\@empty
            \renewcommand\theenumiv{\@arabic\c@enumiv}}%
      \sloppy
      \clubpenalty4000
      \@clubpenalty \clubpenalty
      \widowpenalty4000%
      \sfcode`\.\@m}
     {\def\@noitemerr
       {\@latex@warning{Empty `thebibliography' environment}}%
      \endlist}}%
{\renewenvironment{thebibliography}[1]
     {\section*{\bibname
        \@mkboth{\MakeUppercase\bibname}{\MakeUppercase\bibname}}%
      \list{\@biblabel{\@arabic\c@enumiv}}%
           {\settowidth\labelwidth{\@biblabel{#1}}%
            \leftmargin\labelwidth
            \advance\leftmargin\labelsep
            \itemsep \z@                 % Suppresses vertical separation.
            \@openbib@code
            \usecounter{enumiv}%
            \let\p@enumiv\@empty
            \renewcommand\theenumiv{\@arabic\c@enumiv}}%
      \sloppy
      \clubpenalty4000
      \@clubpenalty \clubpenalty
      \widowpenalty4000%
      \sfcode`\.\@m}
     {\def\@noitemerr
       {\@latex@warning{Empty `thebibliography' environment}}%
      \endlist}}%
\makeatletter
\newcommand{\Exp}{\mathord{\operator@font E}}
\newcommand{\med}{\mathop{\operator@font med}}
\newcommand{\median}{\mathop{\operator@font median}}
\newcommand{\Prob}{\mathord{\operator@font P}}
\newcommand{\sign}{\mathop{\operator@font sign}}
\newcommand{\var}{\mathop{\operator@font var}}
\makeatother
\newtheorem{theorem}{Theorem}[section]
\newtheorem{algorithm}[theorem]{Algorithm}

\newtheorem{corollary}[theorem]{Corollary}

\newtheorem{fact}[theorem]{Fact}
\newtheorem{lemma}[theorem]{Lemma}

\newtheorem{remark}[theorem]{Remark}
\newtheorem{remarks}[theorem]{Remarks}
\begin{document}           % End of preamble and beginning of text.

\maketitle                 % Produces the title.

\begin{abstract}
\noindent
We show that several versions of Floyd and Rivest's improved algorithm
{\sc Select} for finding the $k$th smallest of $n$ elements require at
most $n+\min\{k,n-k\}+O(n^{1/2}\ln^{1/2}n)$ comparisons on average and
with high probability.  This rectifies the analysis of Floyd and Rivest,
and extends it to the case of nondistinct elements.  Encouraging
computational results on large median-finding problems are reported.
\end{abstract}

\begin{quotation}
\end{quotation}

%\begin{quotation}
%\noindent{\bf MSC Subject Classifications.} 68W20, 68W05, 68Q25
%\end{quotation}

%\begin{quotation}
%\noindent{\bf Abbreviated title:} Randomized selection.
%\end{quotation}

%   *** SECTION 1 ***
\section{Introduction}
\label{s:intro}
The {\em selection problem\/} is defined as follows: Given a set
$X:=\{x_j\}_{j=1}^n$ of $n$ elements, a total order $<$ on $X$,
and an integer $1\le k\le n$, find the {\em $k$th smallest\/}
element of $X$, i.e., an element $x$ of $X$ for which there are at
most $k-1$ elements $x_j<x$ and at least $k$ elements $x_j\le x$.
The {\em median\/} of $X$ is the $\lceil n/2\rceil$th smallest
element of $X$.

Selection is one of the fundamental problems in computer science;
see, e.g., the references in \cite{dohaulzw:lbs,dozw:sm,dozw:msr} and
\cite[\S5.3.3]{knu:acpIII2}.  Most references concentrate on the
number of comparisons between pairs of elements made in selection
algorithms.  In the worst case, selection needs at least
$(2+\epsilon)n$ comparisons \cite{dozw:msr}, whereas the algorithm of
\cite{blflprrita:tbs} makes at most $5.43n$, that of \cite{scpapi:fm}
needs $3n+o(n)$, and that in \cite{dozw:sm} takes $2.95n+o(n)$.  In the
average case, for $k\le\lceil n/2\rceil$, at least $n+k-O(1)$
comparisons are necessary \cite{cumu:acs}, whereas Knuth's {\em best
upper bound\/} is $n+k+O(n^{1/2}\ln^{1/2}n)$
\cite[Eq.\ (5.3.3.16)]{knu:acpIII2}.  The classical algorithm {\sc Find}
of \cite{hoa:a65}, also known as quickselect, has an upper bound of
$3.39n+o(n)$ for $k=\lceil n/2\rceil$ in the average case
\cite[Ex.\ 5.2.2--32]{knu:acpIII2}, which improves to $2.75n+o(n)$ for
median-of-3 pivots \cite{gru:mvh,kimapr:ahf}.

The seminal papers \cite{flri:asf,flri:etb} presented three versions
of the algorithm {\sc Select} with very good average case performance,
although their analysis had gaps, as noted in \cite{poriti:eds} and
\cite[Ex.\ 5.3.3--24]{knu:acpIII2}.  Our recent papers
\cite{kiw:rsq,kiw:rst} rectified the analysis of \cite[\S2.2]{flri:etb}
and extended it to the case of nondistinct elements.  Specifically,
we showed that several versions of {\sc Select}, close to those in
\cite[\S2.1]{flri:etb} and \cite{flri:asf}, make at most
%$n+\min\{k,n-k\}+O(n^{2/3}\ln^{1/3}n)$ comparisons on average.
$n+k+O(n^{2/3}\ln^{1/3}n)$ comparisons on average.

This paper concentrates on versions of the improved {\sc Select}
from \cite[\S2.3]{flri:etb}, again correcting its analysis and
extending it to the case of nondistinct elements.  We show that
they make at most $n+k+O(n^{1/2}\ln^{1/2}n)$ comparisons on
average.

Thus, apparently for the first time, Knuth's best upper bound is
attained by an {\em implementable\/} algorithm without
{\em restrictive\/} assumptions.  Specifically, Knuth's scheme
\cite[Ex.\ 5.3.3--24]{knu:acpIII2} is not formulated precisely enough
to qualify as an algorithm, it requires distinct elements in
random order, and its samples are too large for efficient
randomization (since generating a random sample of size
$\lceil n/2\rceil$ takes too much time; cf.\ \S\ref{ss:result}).

We also prove that nonrecursive versions of {\sc Select}, which employ
other linear-time selection routines for small subproblems, require at
most $n+k+O(n^{1/2}\ln^{1/2}n)$ comparisons with high probability; we
couldn't find such results in the literature.  When sorting routines
are used, the bound becomes $n+k+O(n^{1/2}\ln^{3/2}n)$.
% improving upon \cite[Thm 3]{gesi:chb}.

Since our interest is not merely theoretical, a serious effort was made
to implement the various versions efficiently and to test them in
practice.  Our tests on the median-finding examples of \cite{val:iss}
show that the improved {\sc Select} is as fast as the ternary version
of \cite{kiw:rst}, although a bit slower than the quintary version of
\cite{kiw:rsq}.  All these versions perform very well in
terms of the number of comparisons made on large inputs, the average
numbers being about $1.6n$ for $n=1{\rm M}$, and as small as $1.53n$ for
$n=16{\rm M}$.  Since the lower bound is $1.5n$, little room for
improvement remains.  Of course, future work should assess more fully
the relative merits of these versions, but clearly the improved
{\sc Select} may compete with other methods in both theory and practice.

The paper is organized as follows.  A simplified version of {\sc Select}
that ignores some roundings is introduced in \S\ref{s:alg}, and its
basic features are analyzed in \S\ref{s:analysis}.  The average
performance of {\sc Select} and its practical rounded versions is
studied in \S\ref{s:averrec}.  High probability bounds for nonrecursive
versions are derived in \S\ref{s:analnonrec}.
% The use of the
%partitioning schemes of \cite{kiw:rsq} is discussed in \S\ref{s:impl}.
Finally, our computational results are reported in \S\ref{s:exp}.
%The Appendix contains proofs of certain technical results.
%Finally, we have a conclusion section.

Our notation is fairly standard.
$|A|$ denotes the cardinality of a set $A$.
In a given probability space, $\Prob$ is the probability measure,
and $\Exp$ is the mean-value operator.
%and $\Prob[\cdot|{\cal E}]$ is the
%probability conditioned on an event ${\cal E}$; the
%complement of ${\cal E}$ is denoted by ${\cal E}'$.
%$\BbbR_+$ are the positive reals.
%
%   *** SECTION 2 ***
\section{The algorithm {\sc Select}}
\label{s:alg}
We first recall that the standard version of {\sc Select} proceeds as
follows.  By solving two pivot selection subproblems over a random
sample $S$ from $X$, two elements $u$ and $v$ almost sure to be just
below and above the $k$th are found.  The remaining elements are
compared with $u$ and $v$ to derive a reduced selection problem on the
elements between $u$ and $v$ that is solved recursively.  In general,
%the size of the reduced problem (and hence its cost) becomes smaller
the size of the reduced problem (and hence its cost) diminishes
when a larger sample is used, but then the cost of pivot selection
grows.  To balance these costs, the standard version employs a
relatively small sample.  In contrast, the improved version uses a
much larger ``final'' sample $S$, but $u$ and $v$ are selected
iteratively by using samples from $S$.  More specifically, let
$S_1\subset\cdots\subset S_{\bar l}\subset S_{\bar l+1}=X$ be a nested
series of random samples from $X$.  For each sample $S_l$, two pivots
$u_l$ and $v_l$ are found such that $u_l\le x_k^*\le v_l$ with high
probability, where $x_k^*$ is the $k$th element of $X$.  In particular,
$u_l=x_k^*=v_l$ when $S_l=X$.  For $l\le\bar l$, the positions of
$u_{l+1}$ and $v_{l+1}$ in $S_{l+1}$ are chosen so that
$u_l\le u_{l+1}\le v_{l+1}\le v_l$ with high probability, and hence
$u_l$ and $v_l$ can be used to bound the search for $u_{l+1}$ and
$v_{l+1}$.

For clarity, we first describe {\sc Select} in detail without some
integer round-ups in sample sizes, etc.; more practical versions are
postponed till \S\ref{ss:round}.
%
%   *** ALGORITHM 2.1 ***
\begin{algorithm}
\label{alg:sel}
\rm
\hfil\newline\noindent{\bf {\sc Select}$(X,k)$}
(Selects the $k$th smallest element of $X$, with $1\le k\le n:=|X|$)
\medbreak\noindent{\bf Step 1} ({\em Initiation\/}).
If $n=1$, return $x_1$.
Choose parameters $\alpha\in(0,1/2]$, $s_1:=n^\alpha$,
$r>1$, $\kappa:=1/r$, $\beta\ge\frac14(1-\kappa)^{-2}$, and
$\bar l$ such that $n=r^{2\bar l}s_1$.  Set $\theta:=k/n$ and $l:=1$.
\medbreak\noindent{\bf Step 2} ({\em Initial sample selection\/}).
Draw a random sample $S_1$ of size $s_1$ from $X$.  Set
\begin{equation}
g_l:=\left\{\begin{array}{ll}
\left(\,\beta s_l\ln n\,\right)^{1/2}&
\mbox{if}\ l\le\bar l,\\
0&\mbox{if}\ l=\bar l+1,
\end{array}\right.
\label{gl}
\end{equation}
\begin{equation}
i_u^l:=\max\left\{\,\lceil\theta s_l-g_l\rceil,1\,\right\}
\quad\mbox{and}\quad
i_v^l:=\min\left\{\,\lceil\theta s_l+g_l\rceil,s_l\,\right\},
\label{iuivl}
\end{equation}
$u_1:=\mbox{\sc Select}(S_1,i_u^1)$ and
$v_1:=\mbox{\sc Select}(S_1,i_v^1)$ by using {\sc Select} recursively.
\medbreak\noindent{\bf Step 3} ({\em Sample selection\/}).
Draw a random sample $S_{l+1}$ of size $s_{l+1}:=r^2s_l$ from $X$
such that $S_l\subset S_{l+1}$.  (Here $s_{l+1}-s_l$ elements of
$X\setminus S_l$ are picked randomly.)
\medbreak\noindent{\bf Step 4} ({\em Partitioning\/}).
By comparing each element $x$ of $S_{l+1}\setminus S_l$ to $u:=u_l$ and
$v:=v_l$, partition $S_{l+1}$ into $L:=\{x\in S_{l+1}:x<u\}$,
$U:=\{x\in S_{l+1}:x=u\}$, $M:=\{x\in S_{l+1}:u<x<v\}$,
$V:=\{x\in S_{l+1}:x=v\}$, $R:=\{x\in S_{l+1}:v<x\}$.
If $\theta<1/2$, $x$ is compared to $v$ first, and to $u$ only if $x<v$.
If $\theta\ge1/2$, the order of the comparisons is reversed.
\medbreak\noindent{\bf Step 5} ({\em Pivot selection\/}).
(a) Set $g_{l+1}$, $i_u^+:=i_u^{l+1}$ and $i_v^+:=i_v^{l+1}$ via
\eqref{gl}--\eqref{iuivl}.  (Here we wish to find $u_{l+1}$ and
$v_{l+1}$ as the $i_u^+$th and $i_v^+$th smallest elements of
$S_{l+1}$.)\\
(b) If $|L|<i_u^+\le|L\cup U|$, set $u_{l+1}:=u$; else if
$|L\cup U\cup M|<i_u^+\le s_{l+1}-|R|$, set $u_{l+1}:=v$; else set
$u_{l+1}:=\mbox{\sc Select}(\hat S_u,\hat\imath_u^+)$, where $\hat S_u$
and $\hat\imath_u^+$ are determined as follows.
If $i_u^+\le|L|$, set $\hat S_u:=L$ and $\hat\imath_u^+:=i_u^+$; else
if $s_{l+1}-|R|<i_u^+$, set $\hat S_u:=R$ and
$\hat\imath_u^+:=i_u^+-s_{l+1}+|R|$; else set $\hat S_u:=M$ and
$\hat\imath_u^+:=i_u^+-|L\cup U|$.\\
(c) Find $v_{l+1}$, and possibly $\hat S_v$ and $\hat\imath_v^+$, as in
(b) with $i_u^+$ replaced by $i_v^+$ and $u_{l+1}$ by $v_{l+1}$.
\medbreak\noindent{\bf Step 6} ({\em Loop\/}).
If $s_{l+1}=n$, return $u_{l+1}$.  Otherwise, increase $l$ by $1$ and
go to Step 3.
\end{algorithm}

A few remarks on the algorithm are in order.
%
%   *** REMARKS 2.2 ***
\begin{remarks}
\label{r:sel}
\rm
(a)
The correctness and finiteness of {\sc Select} stem by induction from
the following observations.  At Step 2, $|S_1|<|X|$.
At Step 5, $\hat S_u$ and $\hat\imath_u^+$ are chosen so that
the $i_u^+$th smallest element of $S_{l+1}$ is the $\hat\imath_u^+$th
smallest element of $\hat S_u$, and $|\hat S_u|<s_{l+1}$ (since
$u,v\not\in\hat S_u$); similarly for $\hat S_v$ and $\hat\imath_v^+$.
The final loop with $l=\bar l$ has $S_{l+1}=X$, $g_{l+1}=0$ and
$i_u^+=\theta n=k$, so $u_{l+1}=v_{l+1}$ is the desired element.
\par(b)
After Step 5 the position of each element of $S_{l+1}$ relative to
$u_{l+1}$ and $v_{l+1}$ is known.  Hence Step 4 need only compare $u$
and $v$ with the elements of $S_{l+1}\setminus S_l$ (e.g., via one of
the quintary partitioning schemes of \cite[\S6]{kiw:rsq}).
%Further, when Step 3 finds $u_1$, we may remove from $S_1$ its first
%$i_u^1$ smallest elements before extracting $v_1$.
\par(c)
The following elementary property is needed in \S\ref{ss:nonround}.
The maximum number of comparisons taken by {\sc Select} on any input of
size $n$ is finite, for each $n$ (because the recursive calls of
Steps 2 and 5 deal with proper subsets of $X$).
\end{remarks}
%
%   *** SECTION 3 ***
\section{Preliminary analysis}
\label{s:analysis}
In this section we analyze general features of sampling used by
{\sc Select}.
%
%   *** SUBSECTION 3.1 ***
\subsection{Sampling deviations and expectation bounds}
\label{ss:sampleexp}
Our analysis hinges on the following bound on the tail of the
hypergeometric distribution established in \cite{hoe:pis} and
rederived shortly in \cite{chv:thd}.
%
%   *** FACT 3.1 ***
\begin{fact}
\label{f:balls2}
Let\/ $s$ balls be chosen uniformly at random from a set of\/ $s_+$
balls, of which $\rho$ are red, and\/ $\rho'$ be the random variable
representing the number of red balls drawn.  Let\/ $p:=\rho/s_+$.  Then
\begin{equation}
\Prob\left[\,\rho'\ge ps+g\,\right]\le e^{-2g^2\!/s}\quad\forall g\ge0.
\label{Pexpg}
\end{equation}
\end{fact}

We shall also need a simple version of the (left) Chebyshev inequality
\cite[\S2.4.2]{kor:hpt}.
%
%   *** FACT 3.2 ***
\begin{fact}
\label{f:Ebound}
Let\/ $\eta$ be a nonnegative random variable such that\/
$\Prob[\eta\le\zeta]=1$ for some constant\/ $\zeta$.  Then\/
$\Exp\eta\le t+\zeta\Prob[\eta>t]$
for all nonnegative real numbers\/ $t$.
\end{fact}
%
%   *** SUBSECTION 3.2 ***
\subsection{Sample ranks and partitioning efficiency}
\label{ss:samplerank}
In this subsection we analyze in detail a fixed iteration $l$ of
{\sc Select}.

For simpler notation, we drop $l$ from the subscripts and superscripts
and replace $l+1$ by $+$.  Thus let $y_1^*\le\ldots y_s^*$ and
$z_1^*\le\ldots z_{s_+}^*$ denote the sorted elements of the samples
$S$ and $S_+$, so that $u=y_{i_u}^*$, $v=y_{i_v}^*$, $u_+=z_{i_u^+}^*$
and $v_+=z_{i_v^+}^*$, where
\begin{equation}
i_u:=\max\left\{\,\lceil\theta s-g\rceil,1\,\right\}
\quad\mbox{and}\quad
i_v:=\min\left\{\,\lceil\theta s+g\rceil,s\,\right\},
\label{iuiv}
\end{equation}
\begin{equation}
\ \ i_u^+:=\max\left\{\,\lceil\theta s_+-g_+\rceil,1\,\right\}
\quad\mbox{and}\quad
i_v^+:=\min\left\{\,\lceil\theta s_++g_+\rceil,s_+\,\right\}.
\label{iuiv+}
\end{equation}
%$i_u:=\max\{\lceil\theta s-g\rceil,1\}$,
%$i_v:=\min\{\lceil\theta s+g\rceil,s\}$,
%$i_u^+:=\max\{\lceil\theta s_+-g_+\rceil,1\}$,
%$i_v^+:=\min\{\lceil\theta s_++g_+\rceil,s_+\}$.
This notation facilitates showing that $u\le u_+\le v_+\le v$ with
high probability.  To deduce that the number of elements between $u$ and
$v$ is small enough, let
\begin{equation}
j_u:=\max\left\{\,\lceil\theta s_+-2gs_+/s\rceil,1\,\right\}
\quad\mbox{and}\quad
j_v:=\min\left\{\,\lceil\theta s_++2gs_+/s\rceil,s_+\,\right\}
\label{jujv}
\end{equation}
be bounding indices; we shall see that
$z_{j_u}^*\le u\le v\le z_{j_v}^*$ with high probability.  Our
argument is similar to that of \cite[Lem.\ 3.3]{kiw:rsq} because $S$ may
be regarded as a random sample from $S_+$; the key difference is that
$g_+\ne0$ in \eqref{iuiv+} if $l<\bar l$, in which case $g$ is
replaced by $(1-\kappa)g$ in our probability bounds.  To this end,
note that, since $\kappa:=1/r=(s/s_+)^{1/2}$, \eqref{gl} yields
\begin{equation}
g-g_+s/s_+=\left\{\begin{array}{ll}
(1-\kappa)g & \mbox{if}\ l<\bar l,\\
g & \mbox{otherwise}.
\end{array}\right.
\label{gg+ss+}
\end{equation}
%
%   *** LEMMA 3.3 ***
\begin{lemma}
\label{l:rank}
{\rm(a)}
$\Prob[u_+<u]\le e^{-2(1-\kappa)^2g^2\!/s}$ if\/
$i_u=\lceil\theta s-g\rceil$.
\par\indent\rlap{\rm(b)}\hphantom{\rm(a)}
$\Prob[u<z_{j_u}^*]\le e^{-2g^2\!/s}$.
\par\indent\rlap{\rm(c)}\hphantom{\rm(a)}
$\Prob[v<v_+]\le e^{-2(1-\kappa)^2g^2\!/s}$ if\/
$i_v=\lceil\theta s+g\rceil$.
\par\indent\rlap{\rm(d)}\hphantom{\rm(a)}
$\Prob[z_{j_v}^*<v]\le e^{-2g^2\!/s}$.
\par\indent\rlap{\rm(e)}\hphantom{\rm(a)}
$i_u\ne\lceil\theta s-g\rceil$ iff\/ $\theta\le g/s${\rm;}
$i_v\ne\lceil\theta s+g\rceil$ iff\/ $1<\theta+g/s$.
\end{lemma}
\begin{proof}
(a) If $z_{i_u^+}^*<y_{i_u}^*$, at least $s-i_u+1$ samples satisfy
$y_i\ge z_{\bar\jmath+1}^*$ with
$\bar\jmath:=\max_{z_j^*=z_{i_u^+}^*}j$.
In the setting of Fact \ref{f:balls2}, we have $\rho:=s_+-\bar\jmath$
red elements $z_j\ge z_{\bar\jmath+1}^*$, $ps=s-\bar\jmath s/s_+$ and
$\rho'\ge s-i_u+1$.  Since $i_u=\lceil\theta s-g\rceil<\theta s-g+1$ and
$\bar\jmath\ge i_u^+\ge\theta s_+-g_+$ by \eqref{iuiv+}, we get
$s-i_u+1-ps>\bar\jmath s/s_+-\theta s+g\ge g-g_+s/s_+$; thus
$\rho'\ge ps+(1-\kappa)g$ by \eqref{gg+ss+}.  Hence
$\Prob[u_+<u]\le\Prob[\rho'\ge ps+(1-\kappa)g]$, and
\eqref{Pexpg} yields the conclusion.

(b) If $y_{i_u}^*<z_{j_u}^*$, $i_u$ samples are at most $z_\rho^*$,
where $\rho:=\max_{z_j^*<z_{j_u}^*}j$.  Thus we have $\rho$ red elements
$z_j\le z_\rho^*$, $ps=\rho s/s_+$ and $\rho'\ge i_u$.  Now,
$1\le\rho\le j_u-1$ implies $2\le j_u=\lceil\theta s_+-2gs_+/s\rceil$ by
\eqref{jujv} and thus $j_u<\theta s_+-2gs_+/s+1$, so
$-\rho s/s_+>-\theta s+2g$.  Hence
$i_u-ps-g\ge\theta s-g-\rho s/s_+-g>0$, i.e., $\rho'>ps+g$; invoke
\eqref{Pexpg} as before.

(c) and (d): Argue symmetrically to (a) and (b);
cf.\ \cite[Proof of Lem.\ 3.3]{kiw:rsq}.

(e) Follows immediately from the properties of $\lceil\cdot\rceil$
\cite[\S1.2.4]{knu:acpI3}.
\qed
\end{proof}

We may now estimate the partitioning costs of Step 4.
%
%   *** LEMMA 3.4 ***
\begin{lemma}
\label{l:comp}
Let\/ $c:=c_l$ denote the number of comparisons made at Step\/ $4$.
Then\/
\begin{subequations}
\label{PEcomp}
\begin{equation}
\Prob[\,c<\bar c\,]\ge1-e^{-2g^2\!/s}\quad\mbox{and}\quad
\Exp c\le\bar c+2(s_+-s)e^{-2g^2\!/s}\quad\mbox{with}\quad
\label{PEcomp:a}
\end{equation}
\begin{equation}
\bar c:=\left(\,1+\min\{\,\theta,1-\theta\,\}\,\right)(s_+-s)+
3gs_+/s.
\label{PEcomp:b}
\end{equation}
\end{subequations}
%In general, $c\le2(s_+-s)$.
\end{lemma}
\begin{proof}
Consider the event ${\cal A}:=\{c<\bar c\}$ and its complement
${\cal A}':=\{c\ge\bar c\}$.  If $u=v$ then $c=s_+-s<\bar c$; hence
$\Prob[{\cal A}']=\Prob[{\cal A}'\cap\{u<v\}]$, and we may assume
$u<v$ below.

First, suppose $\theta<1/2$.  Then
$c=s_+-s+|\{z\in S_+\setminus S:z<v\}|$, since $s_+-s$ elements of
$S_+\setminus S$ are compared to $v$ first.  In particular,
$c\le2(s_+-s)$.  If $v\le z_{j_v}^*$, then
$\{z\in S_+:z<v\}\subset\{z\in S_+:z<z_{j_v}^*\}$ gives
$|\{z\in S_+:z<v\}|\le j_v-1<\theta s_++2gs_+/s$ by \eqref{jujv},
whereas $u<v$ implies
$|\{z\in S:z<v\}|\ge|\{z\in S:z\le u\}|\ge i_u\ge\theta s-g$
by \eqref{iuiv}, so
$|\{z\in S_+\setminus S:z<v\}|<\theta(s_+-s)+2gs_+/s+g$ yields
$c<\bar c$.  Thus $u<v\le z_{j_v}^*$ implies ${\cal A}$.  Therefore,
${\cal A}'\cap\{u<v\}$ implies $\{z_{j_v}^*<v\}\cap\{u<v\}$, so
$\Prob[{\cal A}'\cap\{u<v\}]\le\Prob[z_{j_v}^*<v]\le e^{-2g^2\!/s}$
(Lem.\ \ref{l:rank}(d)).  Hence we have \eqref{PEcomp}, since
$\Exp c\le\bar c+2(s_+-s)e^{-2g^2\!/s}$
by Fact \ref{f:Ebound} (with $\eta:=c$, $\zeta:=2(s_+-s)$).

Next, suppose $\theta\ge1/2$.  Now
$c=s_+-s+|\{z\in S_+\setminus S:u<z\}|$, since $s_+-s$ elements of
$S_+\setminus S$ are compared to $u$ first.  If $z_{j_u}^*\le u$, then
$\{z\in S_+:u<z\}\subset\{z\in S_+:z_{j_u}^*<z\}$ gives
$|\{z\in S_+:u<z\}|\le s_+-j_u\le s_+-\theta s_++2gs_+/s$,
whereas $u<v$ implies
$|\{z\in S:u<z\}|\ge|\{z\in S:v\le z\}|\ge s-i_v+1\ge s-\theta s-g+1$,
so $|\{z\in S_+\setminus S:u<z\}|\le(1-\theta)(s_+-s)+2gs_+/s+g-1$
yields $c<\bar c$.  Thus
${\cal A}'\cap\{u<v\}$ implies $\{u<z_{j_u}^*\}\cap\{u<v\}$, so
$\Prob[{\cal A}'\cap\{u<v\}]\le\Prob[u<z_{j_u}^*]\le e^{-2g^2\!/s}$
(Lem.\ \ref{l:rank}(b)), and we get \eqref{PEcomp} as before.
\qed
\end{proof}

The following result will imply that the sets $\hat S_u$ and $\hat S_v$
selected at Step 5 are ``small enough'' with high probability.  Let
$\hat s:=\hat s_l:=|\hat S_u\cup\hat S_v|$; we let
$\hat S_u:=\emptyset$ (or $\hat S_v:=\emptyset$) if Step 5 doesn't use
$\hat S_u$ (or $\hat S_v$), but we don't consider this case explicitly.
%
%   *** LEMMA 3.5 ***
\begin{lemma}
\label{l:PX}
$\Prob\left[\hat s<4gs_+/s\right]\ge1-{\rm P}_{\rm fail}$ and\/
$\hat s<s_+$ always, where\/
\begin{equation}
{\rm P}_{\rm fail}:={\rm P}_{\rm fail}(n):=
2e^{-2g^2\!/s}+2e^{-2(1-\kappa)^2g^2\!/s}=
2n^{-2\beta}+2n^{-2(1-\kappa)^2\beta}\le
4n^{-2(1-\kappa)^2\beta}.
\label{Pfail}
\end{equation}
\end{lemma}
\begin{proof}
First, consider the {\em middle\/} case of $i_u=\lceil\theta s-g\rceil$
and $i_v=\lceil\theta s+g\rceil$.  Let ${\cal E}$ denote the event
$z_{j_u}^*\le u\le u_+\le v_+\le v\le z_{j_v}^*$.  By
Lem.\ \ref{l:rank} and the Boole-Benferroni inequality, its complement
${\cal E}'$ has $\Prob[{\cal E}']\le{\rm P}_{\rm fail}$, so
$\Prob[{\cal E}]\ge1-{\rm P}_{\rm fail}$.  By the rules of Steps 4--5,
$u\le u_+\le v_+\le v$ implies $\hat S_u\cup\hat S_v\subset M$, whereas
$z_{j_u}^*\le u\le v\le z_{j_v}^*$ yields $\hat s\le j_v-j_u+1-2$;
since $j_v<\theta s_++2gs_+/s+1$ and $j_u\ge\theta s_+-2gs_+/s$ by
\eqref{jujv}, we get $\hat s<4gs_+/s$.  Hence
$\Prob[\hat s<4gs_+/s]\ge\Prob[{\cal E}]$.  Then \eqref{Pfail} follows
from \eqref{gl} and the fact $\kappa\in(0,1)$.

Next, consider the {\em left\/} case of
$i_u\ne\lceil\theta s-g\rceil$, i.e., $\theta\le g/s$
(Lem.\ \ref{l:rank}(e)).  If $i_v\ne\lceil\theta s+g\rceil$, then
$1<\theta+g/s$ (Lem.\ \ref{l:rank}(e)) gives $\hat s<s_+<2gs_+/s$.
For $i_v=\lceil\theta s+g\rceil$,
$\Prob[v_+\le v\le z_{j_v}^*]\ge1-\frac12{\rm P}_{\rm fail}$
by Lem.\ \ref{l:rank}(c,d).  Now, $v_+\le v$ implies
$\hat S_u\cup\hat S_v\subset L\cup M$, whereas $v\le z_{j_v}^*$ gives
$\hat s\le j_v-1<\theta s_++2gs_+/s\le3gs_+/s$; hence
$\Prob[\hat s<4gs_+/s]\ge\Prob[v_+\le v\le z_{j_v}^*]$.

Finally, consider the {\em right\/} case of
$i_v\ne\lceil\theta s+g\rceil$, i.e., $1<\theta+g/s$.  If
$i_u\ne\lceil\theta s-g\rceil$ then $\theta\le g/s$ gives
$\hat s<s_+<2gs_+/s$.  For $i_u=\lceil\theta s-g\rceil$, we have
$\Prob[z_{j_u}^*\le u\le u_+]\ge1-\frac12{\rm P}_{\rm fail}$
by Lem.\ \ref{l:rank}(a,b).  Now, $u\le u_+$ implies
$\hat S_u\cup\hat S_v\subset M\cup R$, whereas $z_{j_u}^*\le u$ yields
$\hat s\le s_+-j_u$ with $j_u\ge\theta s_+-2gs_+/s$ and thus
$\hat s<3gs_+/s$, so
$\Prob[\hat s<4gs_+/s]\ge\Prob[z_{j_u}^*\le u\le u_+]$.
\qed
\end{proof}
%
%   *** COROLLARY 3.6 ***
\begin{corollary}
\label{c:PcX}
$\Prob\left[c<\bar c\ \mbox{and}\ \hat s<4gs_+/s\right]\ge
1-{\rm P}_{\rm fail}$.
\end{corollary}
\begin{proof}
If $2g/s\ge1$ then $c\le2(s_+-s)<\bar c$ (cf.\ \eqref{PEcomp:b})
and $\hat s<s_+<4gs_+/s$, so assume $2g/s<1$.  The conclusion
follows from the proofs of Lems.\ \ref{l:comp} and \ref{l:PX}.  We
only note that the left case of $\theta\le g/s$ now has
$i_v=\lceil\theta s+g\rceil$ and $\theta<1/2$.  Similarly, in the
right case of $1<\theta+g/s$, we have $i_u=\lceil\theta s-g\rceil$ and
$\theta\ge1/2$, since $g/s<1/2$.
\qed
\end{proof}
%
%   *** REMARK 3.7 ***
\begin{remark}
\label{r:PcX}
\rm
Suppose for $l<\bar l$, Step 5 resets $i_u^+:=i_v^+$ if
$\theta\le g_{l+1}/s_{l+1}$, or $i_v^+:=i_u^+$ if
$1<\theta+g_{l+1}/s_{l+1}$, finding a single pivot $u_+=v_+$ in these
cases.  The preceding results remain valid for this modification (which
corresponds to using $u:=v$ if $\theta\le g/s$, or $v:=u$ if
$1<\theta+g/s$).
%Note that $\theta\le g/s$ implies $j_u=1$ and $z_{j_u}^*\le u$,
%whereas $1<\theta+g/s$ yields $j_v=s_+$ and $v\le z_{j_v}^*$.
Similarly, Step 2 may reset $i_u^1:=i_v^1$ if $\theta\le g_1/s_1$, or
$i_v^1:=i_u^1$ if $1<\theta+g_1/s_1$.
\end{remark}
%
%   *** SECTION 4 ***
\section{Average performance of the recursive version}
\label{s:averrec}
%
%   *** SUBSECTION 4.1 ***
\subsection{Analysis of the nonrounded version}
\label{ss:nonround}
In this section we analyze the average performance of {\sc Select},
starting with the ``nonrounded'' version of Algorithm \ref{alg:sel};
more practical versions are discussed in \S\ref{ss:round}.
%
%   *** THEOREM 4.1 ***
\begin{theorem}
\label{t:sel}
Let\/ $C_{nk}$ denote the expected number of comparisons made by\/
{\sc Select}, and\/ $f(t):=(t\ln t)^{1/2}$ for\/ $t\ge1$.
There exists a positive constant\/ $\gamma$ such that
\begin{equation}
C_{nk}\le n+\min\{\,k,n-k\,\}+\gamma f(n)
\quad\mbox{for all\/}\ 1\le k\le n.
\label{Cnk}
\end{equation}
\end{theorem}
\begin{proof}
We need a few preliminary facts.
The function $\phi(t):=f(t)/t=(\ln t/t)^{1/2}$ decreases to $0$ on
$[e,\infty)$, whereas $f(t)$ grows to infinity on $[2,\infty)$.
The key {\em bounding property\/} is $f(t)=\phi(t)t\le\phi(\hat t)t$
for all $t\ge\hat t\ge e$.  Pick $\bar n\ge2$ large enough so that
$s_1\ge e$, $4r^2g_1\ge e$, $n^\alpha+1\le f(n)$ and $n\le r^2s_1$ for
all $n\ge\bar n$.  Using $\alpha\in(0,1/2]$ and the bounding property,
we have
\begin{equation}
s_1\le f(n)\quad\mbox{and}\quad f(s_1)\le\phi(s_1)f(n).
\label{s1fs1}
\end{equation}
By \eqref{Pfail} and our assumption $\beta\ge\frac14(1-\kappa)^{-2}$,
we have $n{\rm P}_{\rm fail}(n)=o(f(n))$; more precisely,
\begin{equation}
n{\rm P}_{\rm fail}(n)\le4n^{1-2(1-\kappa)^2\beta}=
4f(n)n^{1/2-2(1-\kappa)^2\beta}\ln^{-1/2}n.
\label{nPfail}
\end{equation}
Using the monotonicity of $\phi$, we may increase $\bar n$ if necessary
to get for all $n\ge\bar n$
\begin{equation}
\phi(s_1)+4\frac{2r^2-r}{r-1}\beta^{1/2}\phi(4r^2g_1)+
4\frac{2r^2-1}{r^2-1}\phi(r^2s_1)n^{1/2-2(1-\kappa)^2\beta}
\ln^{-1/2}n\le0.475,
\label{0.475reqint}
\end{equation}
since each term above goes to $0$ as $n$ increases to $\infty$.
By Rem.\ \ref{r:sel}(c), there is $\gamma$ such that \eqref{Cnk}
holds for all $n\le\bar n$; increasing $\gamma$ if necessary, we have
for all $n\ge\bar n$
\begin{equation}
3+15\frac{2r^2-r}{r-1}\beta^{1/2}+
4\frac{6.5r^2-3.5}{r^2-1}n^{1/2-2(1-\kappa)^2\beta}\ln^{-1/2}n\le
0.05\gamma.
\label{0.05gamint}
\end{equation}

Let $n'\ge\bar n$.  Assuming \eqref{Cnk} holds for all $n\le n'$,
for induction let $n=n'+1$.

Since $s_1<n$, by our hypothesis the cost of selecting $u_1$ and $v_1$
at Step 2 is at most
\begin{equation}
C_{s_1i_u^1}+C_{s_1i_v^1}\le3s_1+2\gamma f(s_1).
\label{Cs1iu1}
\end{equation}
Similarly, the cost of selecting $u_{l+1}$ and $v_{l+1}$ at Step 5 is
at most $3\hat s_l+2\gamma f(\hat s_l)$, where $\hat s_l<s_{l+1}$ and
$\Prob[\hat s_l\ge 4g_ls_{l+1}/s_l]\le{\rm P}_{\rm fail}$ by
Lem.\ \ref{l:PX}.  Hence (cf.\ Fact \ref{f:Ebound} with
$\eta:=3\hat s_l+2\gamma f(\hat s_l)$)
\begin{equation}
\Exp\left[\,3\hat s_l+2\gamma f(\hat s_l)\,\right]\le
12g_ls_{l+1}/s_l+2\gamma f(4g_ls_{l+1}/s_l)+
\left[\,3s_{l+1}+2\gamma f(s_{l+1})\,\right]{\rm P}_{\rm fail},
\quad l=1\colon\bar l.
\label{E3hats}
\end{equation}
For $\bar\theta:=\min\{\theta,1-\theta\}$, the partitioning cost of
Step 4 is estimated by \eqref{PEcomp} as
\begin{equation}
\Exp c_l\le\left(\,1+\bar\theta\,\right)(s_{l+1}-s_l)+3g_ls_{l+1}/s_l+
{\textstyle\frac12}(s_{l+1}-s_l){\rm P}_{\rm fail},
\quad l=1\colon\bar l.
\label{Ecl}
\end{equation}

Adding the costs \eqref{Cs1iu1}--\eqref{Ecl} and using
$s_{\bar l+1}=n$, we get
\begin{subequations}
\label{Cnk1theta}
\begin{eqnarray}
\label{Cnk1theta:a}
C_{nk}&\le&\left(\,1+\bar\theta\,\right)(n-s_1)+\left[\,3s_1+
15\sum_{l=1}^{\bar l}g_ls_{l+1}/s_l+
{\textstyle\frac12}{\rm P}_{\rm fail}(n-s_1)+
3{\rm P}_{\rm fail}\sum_{l=1}^{\bar l}s_{l+1}\,\right]\quad\qquad\\
\label{Cnk1theta:b}
&&\quad{}+2\gamma\left[\,f(s_1)+
\sum_{l=1}^{\bar l}f(4g_ls_{l+1}/s_l)+
{\rm P}_{\rm fail}\sum_{l=1}^{\bar l}f(s_{l+1})\,\right].
\end{eqnarray}
\end{subequations}
Since $\theta:=k/n$, the first term on the right side above is at
most $n+\min\{k,n-k\}$.  Next, for $d:=(\beta\ln n)^{1/2}$, \eqref{gl}
yields $g_ls_{l+1}/s_l=ds_{l+1}/s_l^{1/2}$ for $l\le\bar l$.  Since
$s_l=r^{2(l-1)}s_1$ for $l\le\bar l$, and $n>r^{2(\bar l-1)}s_1$
implies $r^{\bar l-1}<(n/s_1)^{1/2}$, we obtain
$$%\begin{equation}
\sum_{l=1}^{\bar l-1}g_ls_{l+1}/s_l=
\sum_{l=1}^{\bar l-1}dr^{l+1}s_1^{1/2}=
dr^2s_1^{1/2}\frac{r^{\bar l-1}-1}{r-1}<
\beta^{1/2}f(n)\frac{r^2}{r-1}.
%\label{sumgl-1int}
$$%\end{equation}
But $g_{\bar l}s_{\bar l+1}/s_{\bar l}=dn/s_{\bar l}^{1/2}=
\beta^{1/2}f(n)(n/s_{\bar l})^{1/2}$ and $n\le r^2s_{\bar l}$
imply $g_{\bar l}s_{\bar l+1}/s_{\bar l}\le\beta^{1/2}f(n)r$, so
\begin{equation}
\sum_{l=1}^{\bar l}g_ls_{l+1}/s_l<
\beta^{1/2}f(n)\left(\,\frac{r^2}{r-1}+r\,\right)=
\beta^{1/2}f(n)\frac{2r^2-r}{r-1}.
\label{sumglsl+1sl}
\end{equation}
Similarly, using $s_{l+1}=r^{2l}s_1$ for $l<\bar l$, $s_{\bar l+1}=n$
and $r^{2(\bar l-1)}<n/s_1$, we get
\begin{equation}
\sum_{l=1}^{\bar l}s_{l+1}=
s_1\sum_{l=1}^{\bar l-1}r^{2l}+n=
r^2s_1\frac{r^{2(\bar l-1)}-1}{r^2-1}+n<
r^2\frac{n-s_1}{r^2-1}+n<\frac{2r^2-1}{r^2-1}n.
\label{sumsl+1int}
\end{equation}
Plugging \eqref{s1fs1}, \eqref{nPfail}, \eqref{sumglsl+1sl} and
\eqref{sumsl+1int} into \eqref{Cnk1theta:a}, we see that the bracketed
term is at most $0.05\gamma f(n)$ thanks to \eqref{0.05gamint}.
Next, for $l<\bar l$ we have
$4g_ls_{l+1}/s_l\ge4r^2g_1$ (cf.\ \eqref{gl}), whereas
$g_{\bar l}s_{\bar l+1}/s_{\bar l}\le\beta^{1/2}f(n)r$ with
$4\beta^{1/2}f(n)r\ge4r^2g_1$ from $n\ge r^2s_1$; therefore, we may
use the bounding property and argue as for \eqref{sumglsl+1sl} to get
%\begin{eqnarray}
%\sum_{l=1}^{\bar l}f(4g_ls_{l+1}/s_l)&\le&
%\phi(4r^2g_1)4\left(\,\sum_{l=1}^{\bar l-1}g_ls_{l+1}/s_l+
%\beta^{1/2}f(n)r\,\right)\nonumber\\
%&<&
%4\frac{2r^2-r}{r-1}\beta^{1/2}\phi(4r^2g_1)f(n).
%\label{sumf4glsl+1sl}
%\end{eqnarray}
\begin{equation}
\sum_{l=1}^{\bar l}f(4g_ls_{l+1}/s_l)\le
\phi(4r^2g_1)4\left(\,\sum_{l=1}^{\bar l-1}g_ls_{l+1}/s_l+
\beta^{1/2}f(n)r\,\right)<
4\frac{2r^2-r}{r-1}\beta^{1/2}\phi(4r^2g_1)f(n).
\label{sumf4glsl+1sl}
\end{equation}
Similarly, $s_{l+1}=r^{2l}s_1\ge r^2s_1$ for $l<\bar l$ and
$s_{\bar l+1}=n\ge r^2s_1$ together with \eqref{sumsl+1int}
imply
\begin{equation}
\sum_{l=1}^{\bar l}f(s_{l+1})\le
\phi(r^2s_1)\sum_{l=1}^{\bar l}s_{l+1}<
\frac{2r^2-1}{r^2-1}\phi(r^2s_1)n.
\label{sumfsl+1int}
\end{equation}
Now, plugging \eqref{s1fs1}, \eqref{sumf4glsl+1sl} and
\eqref{sumfsl+1int} combined with \eqref{nPfail} into
\eqref{Cnk1theta:b}, we deduce that \eqref{Cnk1theta:b} is at most
$0.95\gamma f(n)$ due to \eqref{0.475reqint}; thus \eqref{Cnk} holds
as required.
\qed
\end{proof}
%
%   *** SUBSECTION 4.2 ***
\subsection{Analysis of rounded versions}
\label{ss:round}
We now consider more realistic parameter choices for {\sc Select}.

Fixing $\alpha\in(0,1/2]$, $r>1$ such that $r^2$ is integer,
$\kappa:=1/r$, $\beta\ge\frac14(1-\kappa)^{-2}$, suppose % for $n>1$
Steps 1 and 3 set
\begin{equation}
s_1:=\min\left\{\,\lceil n^\alpha\rceil,n-1\,\right\},
\label{s1int}
\end{equation}
\begin{equation}
\bar l:=\min\left\{\,l:r^{2l}s_1\ge n\,\right\}=
\left\lceil\ln(n/s_1)/\ln r^2\right\rceil,
\label{lbarint}
\end{equation}
\begin{equation}
s_{l+1}:=\min\left\{\,r^{2l}s_1,n\,\right\}=
\min\left\{\,r^2s_l,n\,\right\}.
\label{sl+1int}
\end{equation}
Note that \eqref{s1int}--\eqref{sl+1int} yield
$s_{l+1}=r^{2l}s_1$ if $l<\bar l$,
$s_{\bar l+1}=n>r^{2(\bar l-1)}s_1$.  It is easy to see that the
proof of Theorem \ref{t:sel} covers this modification.

The final iteration $\bar l$ doesn't need sampling, since
$S_{\bar l+1}=X$.  Hence, to reduce the sampling costs, we may wish to
ensure that $s_{\bar l}$, the number of sampled elements, is at most
a fixed fraction $\bar\eta\in(1/r^2,1]$ of $n$ when $n$ is large.  To
this end, suppose that for
\begin{equation}
n\ge\max\left\{\,[\,r^2\!/(\bar\eta r^2-1)\,]^{1/\alpha},
3\,\right\}\quad\mbox{with}\quad\bar\eta\in(1/r^2,1],
\label{n>=r2}
\end{equation}
we replace \eqref{s1int}--\eqref{lbarint} by
\begin{equation}
\bar l:=\min\left\{\,l:r^{2l}n^\alpha\ge n\,\right\}=
%\left\lceil{\textstyle\frac12}(1-\alpha)\ln n/\ln r\right\rceil,
\left\lceil(1-\alpha)\ln n/\ln r^2\right\rceil,
\label{lbarS}
\end{equation}
\begin{equation}
s_1:=\left\lceil\,n/r^{2\bar l}\,\right\rceil.
\label{s1S}
\end{equation}
Then $n^\alpha\!/r^2\le s_1\le\lceil n^\alpha\rceil<n$ replaces
\eqref{s1int}, \eqref{lbarint} remains true and
\begin{equation}
s_{\bar l}<\eta n\quad\mbox{with}\quad
\eta:=r^{-2}+n^{-\alpha}\le\bar\eta.
\label{sbarletaint}
\end{equation}
Indeed, $n\le r^{2\bar l}n^\alpha<r^2n$ implies
$n^\alpha\!/r^2\le s_1\le\lceil n^\alpha\rceil$; since
$n^\alpha\le n^{1/2}\le n-1$ for $n\ge3$, we have
$\lceil n^\alpha\rceil<n$.
Next, $n/r^{2\bar l}>n^\alpha\!/r^2=1/(\eta r^2-1)$ yields
$\eta n/r^{2(\bar l-1)}>n/r^{2\bar l}+1>s_1$; thus
$\eta n>r^{2(\bar l-1)}s_1$.
But $n^\alpha\ge r^2\!/(\bar\eta r^2-1)$ implies $\eta\le\bar\eta\le1$,
so $r^{2(\bar l-1)}s_1<n$, \eqref{lbarint} holds and \eqref{sl+1int}
gives $s_{\bar l}<\eta n$.  In effect, Theorem \ref{t:sel} holds
for this modification.
%
%   *** SUBSECTION 4.3 ***
\subsection{Using smaller rank gaps}
\label{ss:smallgap}
Although the gaps $g_l$ of \eqref{gl} give useful high probability
bounds (cf.\ \S\ref{s:analnonrec}), in practice the average
performance on small problems improves for the smaller gaps
\begin{equation}
g_l:=\left(\,\beta s_l\ln s_l\,\right)^{1/2}
\quad\mbox{for}\ l\le\bar l.
\label{glsmall}
\end{equation}
Assuming $\beta>\frac14(1-\kappa)^{-2}$, we now sketch briefly
how to extend the previous results.
First, $\psi(s):=[1-\kappa(1+\ln r^2\!/\ln s)^{1/2}]^2$ replaces
$(1-\kappa)^2$ in the relations of \S\ref{ss:samplerank}, and
\eqref{Pfail} becomes
\begin{equation}
{\rm P}_{\rm fail}:={\rm P}_{\rm fail}(s):=
2e^{-2g^2\!/s}+2e^{-2\psi(s)g^2\!/s}=
2s^{-2\beta}+2s^{-2\beta\psi(s)}\le
4s^{-2\beta\psi(s)}.
\label{Pfailsmall}
\end{equation}
For $\bar n$ such that $2\beta\psi(s_1)\ge1/2$ for all
$n\ge\bar n$, \eqref{E3hats}--\eqref{Ecl} now involve
${\rm P}_{\rm fail}(s_l)\le4s_l^{-1/2}$, so \eqref{Cnk1theta} is
modified accordingly, whereas \eqref{sumsl+1int} and
\eqref{sumfsl+1int} are replaced by
\begin{equation}
\sum_{l=1}^{\bar l}s_{l+1}{\rm P}_{\rm fail}(s_l)/4\le
\sum_{l=1}^{\bar l-1}r^2s_l^{1/2}+n^{1/2}r<
r^2\frac{n^{1/2}-s_1^{1/2}}{r-1}+n^{1/2}r<
\frac{2r^2-1}{r-1}n^{1/2},
\label{sumsl+1small}
\end{equation}
\begin{equation}
\sum_{l=1}^{\bar l}f(s_{l+1}){\rm P}_{\rm fail}(s_l)\le
\phi(r^2s_1)\sum_{l=1}^{\bar l}s_{l+1}{\rm P}_{\rm fail}(s_l)<
4\frac{2r^2-1}{r-1}\phi(r^2s_1)n^{1/2}.
\label{sumfsl+1small}
\end{equation}
Modify the third terms of \eqref{0.475reqint}--\eqref{0.05gamint} to
complete the proof of Theorem \ref{t:sel} as before.
%
%   *** SUBSECTION 4.4 ***
\subsection{Handling small subfiles}
\label{ss:subfile}
Since the sampling efficiency decreases when $X$ shrinks, consider the
following modification.  For a fixed cut-off parameter
$n_{\rm cut}\ge1$, let sSelect$(X,k)$ be a ``small-select'' routine that
finds the $k$th smallest element of $X$ in at most $C_{\rm cut}<\infty$
comparisons when $|X|\le n_{\rm cut}$ (even bubble sort will do).  Then
{\sc Select} is modified to start with the following
\medbreak\noindent{\bf Step 0} ({\em Small file case\/}).
If $n:=|X|\le n_{\rm cut}$, return sSelect$(X,k)$.

Our preceding results remain valid for this modification.  In fact it
suffices if $C_{\rm cut}$ bounds the {\em expected\/} number of
comparisons of sSelect$(X,k)$ for $n\le n_{\rm cut}$.  For instance,
\eqref{Cnk} holds for $n\le n_{\rm cut}$ and $\gamma\ge C_{\rm cut}$,
and by induction as in Rem.\ \ref{r:sel}(c) we have $C_{nk}<\infty$
for all $n$, which suffices for the proof of Theorem \ref{t:sel}.

%Another advantage is that even small $n_{\rm cut}$ limits the stack
%space for recursion.
%
%   *** SECTION 5 ***
\section{Analysis of nonrecursive versions}
\label{s:analnonrec}
Consider a {\em nonrecursive version\/} of {\sc Select} in which Steps
2 and 5, instead of {\sc Select}, employ a linear-time routine (e.g.,
{\sc Pick} \cite{blflprrita:tbs}) that finds the $i$th smallest of $m$
elements in at most $\gamma_Pm$ comparisons for some constant
$\gamma_P>2$.
%
%   *** THEOREM 5.1 ***
\begin{theorem}
\label{t:selnonrec}
Let\/ $c_{nk}$ denote the number of comparisons made by the nonrecursive
version of\/ {\sc Select}, using \eqref{s1int}--\eqref{sl+1int}.
Then for\/ $n\ge6$, we have
\begin{subequations}
\label{cnk}
\begin{equation}
\Prob\left[\,c_{nk}\le n+\min\{\,k,n-k\,\}+\hat\gamma_Pf(n)\,\right]\ge
1-\bar l{\rm P}_{\rm fail}\quad\mbox{with}
\label{cnk:a}
\end{equation}
\begin{equation}
\hat\gamma_P:=2\gamma_P+\frac{2r^2-r}{r-1}(3+8\gamma_P)\beta^{1/2},
\label{cnk:b}
\end{equation}
\begin{equation}
\bar l{\rm P}_{\rm fail}\le
4\left\lceil\,(1-\alpha)\ln n/\ln r^2\,\right\rceil
n^{-2(1-\kappa)^2\beta}.
\label{cnk:c}
\end{equation}
\end{subequations}
In particular, $\bar l{\rm P}_{\rm fail}=o(n^{-1})$ if\/
$\beta>\frac12(1-\kappa)^{-2}$.  Moreover,
\begin{subequations}
\label{Ecnk}
\begin{equation}
\Exp c_{nk}\le n+\min\{\,k,n-k\,\}+\bar\gamma_Pf(n)\quad\mbox{with}
\label{Ecnk:a}
\end{equation}
\begin{equation}
\bar\gamma_P:=\hat\gamma_P+
4\left(\,\frac{2r^2-1}{r^2-1}2\gamma_P+1/2\,\right)
n^{1/2-2(1-\kappa)^2\beta}\ln^{-1/2}n.
\label{Ecnk:b}
\end{equation}
\end{subequations}
In particular, $\bar\gamma_P\le\hat\gamma_P+16\gamma_P+2$ if\/
$\beta\ge\frac14(1-\kappa)^{-2}$.
\end{theorem}
\begin{proof}
The cost of Step 2 is at most $2\gamma_Ps_1$, with
$s_1\le\lceil n^{1/2}\rceil\le f(n)\le n-1$, since $n\ge6$.
For $\bar\theta:=\min\{\theta,1-\theta\}$, the cost of
Steps 4 and 5 at iteration $l$ is at most
\begin{equation}
\bar C_l:=\left(\,1+\bar\theta\,\right)(s_{l+1}-s_l)+3g_ls_{l+1}/s_l+
2\gamma_P\cdot4g_ls_{l+1}/s_l
\label{barCl}
\end{equation}
with probability at least $1-{\rm P}_{\rm fail}$ by \eqref{PEcomp}
and Cor.\ \ref{c:PcX}.  Hence $c_{nk}$ exceeds
$$
\bar C:=2\gamma_Ps_1+\sum_{l=1}^{\bar l}\bar C_l=
2\gamma_Ps_1+\left(\,1+\bar\theta\,\right)(n-s_1)+(3+8\gamma_P)
\sum_{l=1}^{\bar l}g_ls_{l+1}/s_l
$$
with probability at most $\bar l{\rm P}_{\rm fail}$.  But
$\bar C\le n+\min\{k,n-k\}+\hat\gamma_Pf(n)$ by \eqref{sumglsl+1sl}
and \eqref{cnk:b}, so \eqref{cnk:a} follows.  Then \eqref{Pfail}
and \eqref{lbarint} with $s_1\ge n^\alpha$ yield \eqref{cnk:c}.

Similarly, $\Exp c_{nk}\le2\gamma_Ps_1+
\sum_{l=1}^{\bar l}(\Exp c_l+2\gamma_P\Exp\hat s_l)$; bounding these
costs as for \eqref{E3hats}--\eqref{Ecl} via \eqref{nPfail},
\eqref{sumglsl+1sl} and \eqref{sumsl+1int} gives \eqref{Ecnk}.
\qed
\end{proof}
%
%   *** REMARKS 5.2 ***
\begin{remarks}
\label{r:selnonrec}
\rm
(a)
The bound \eqref{Ecnk} holds if Steps 2 and 5 employ a routine (e.g.,
{\sc Find} \cite{hoa:a65}) for which the {\em expected\/} number of
comparisons to find the $i$th smallest of $m$ elements is at most
$\gamma_Pm$ (then $\Exp c_{nk}$ is bounded as before).
\par(b)
Suppose Step 5 returns to Step 2 if $\hat s_l\ge4g_ls_{l+1}/s_l$.
By Cor.\ \ref{c:PcX}, such loops are finite wp $1$, and don't
occur with high probability, for $n$ large enough.
\par(c)
Suppose Steps 2 and 5 simply sort $S$ and $\hat S_u\cup\hat S_v$ by any
algorithm that takes at most $\gamma_Sm\ln m$ comparisons to sort $m$
elements for a constant $\gamma_S$.  Then the cost of Step 2 is at
most $\gamma_Ss_1\ln n$, because $s_1<n$; hence $\gamma_S\ln n$ may
replace $2\gamma_P$ in \eqref{cnk:b}.  Similarly, $\gamma_S\ln n$
replaces $\gamma_P$ in \eqref{barCl} and \eqref{Ecnk:b}, and
$4\gamma_S\ln n$ replaces $8\gamma_P$ in \eqref{cnk:b}.  In other words,
$n^{1/2}\ln^{3/2}n$ replaces $f(n)$ in \eqref{cnk:a} and \eqref{Ecnk:a}
for suitably redefined $\hat\gamma_P$ and $\bar\gamma_P$.
\end{remarks}
%
%   *** SECTION 6 ***
\section{Experimental results}
\label{s:exp}
%
%   *** SUBSECTION 6.1 ***
\subsection{Implemented algorithms}
\label{ss:impl}
An implementation of {\sc Select} was programmed in Fortran 77 and
run on a notebook PC (Pentium 4M 2 GHz, 768 MB RAM) under MS
Windows XP.  The input set $X$ was specified as a double precision
array, and the partitioning schemes of \cite[\S6]{kiw:rsq} were used.
For efficiency, small arrays with $n\le n_{\rm cut}$ were
handled by sSelect (cf.\ \S\ref{ss:subfile}), which
typically required less than $3.5n$ comparisons.  We used
$n_{\rm cut}=600$ as proposed in \cite{flri:asf}, $\alpha=0.5$,
$\beta=0.3$ in \eqref{glsmall}, $r=12$ and $\bar\eta=2/r^2$; future
work should test other parameters.
%
%   *** SUBSECTION 6.2 ***
\subsection{Testing examples}
\label{ss:examp}
As in \cite{kiw:rsq}, we used minor modifications of the input sequences
of \cite{val:iss}:
\begin{description}
\itemsep0pt
\item[random]
A random permutation of the integers $1$ through $n$.
\item[onezero]
A random permutation of $\lceil n/2\rceil$ ones and $\lfloor n/2\rfloor$
zeros.
\item[sorted]
The integers $1$ through $n$ in increasing order.
\item[organpipe]
The integers $(1,2,\ldots,n/2,n/2,\ldots,2,1)$.
\end{description}
For each input sequence, its (lower) median element was selected
for $k:=\lceil n/2\rceil$.  To save space, we only add that the
results for the twofaced, rotated and m3killer sequences of
\cite{kiw:rsq} were similar to those of the random, sorted and
organpipe inputs, respectively.
%
%   *** SUBSECTION 6.3 ***
\subsection{Computational results}
\label{ss:result}
We varied the input size $n$ from $50{,}000$ to $16{,}000{,}000$.  For
the random and onezero sequences, for each input size,
20 instances were randomly generated; for the deterministic
sequences, 20 runs were made to measure the solution time.

The performance of {\sc Select} is summarized in
Table \ref{tab:Selrand},
%
%   *** TABLE 6.1 ***
\begin{table}[t!]
\caption{Performance of {\sc Select} on randomly generated inputs.}
\label{tab:Selrand}
\footnotesize
\begin{center}
\tabcolsep=0.98\tabcolsep
\begin{tabular}{lrrrrrrrrrrrrr}
\hline
Input &\multicolumn{1}{c}{Size}
&\multicolumn{3}{c}{Time $[{\rm msec}]$%
\vphantom{$1^{2^3}$}} % Need more vertical space!
&\multicolumn{3}{c}{Comparisons $[n]$}
&\multicolumn{1}{c}{$\gamma_{\rm avg}$}
&\multicolumn{1}{c}{$L_{\rm avg}$}
&\multicolumn{1}{c}{$P_{\rm avg}$}
&\multicolumn{1}{c}{$N_{\rm avg}$}
&\multicolumn{1}{c}{$p_{\rm avg}$}
&\multicolumn{1}{c}{$s_{\rm avg}$}\\
&\multicolumn{1}{c}{$n$}
&\multicolumn{1}{c}{avg}&\multicolumn{1}{c}{max}&\multicolumn{1}{c}{min}
&\multicolumn{1}{c}{avg}&\multicolumn{1}{c}{max}&\multicolumn{1}{c}{min}
& &\multicolumn{1}{c}{$[n]$}
&\multicolumn{1}{c}{$[\ln n]$}
&\multicolumn{1}{c}{$[\ln n]$} &
&\multicolumn{1}{c}{$[\%n]$}\\
\hline
%dsel30a/dsel30ax
random     &  50K
&    2&   10&    0& 1.89& 2.05& 1.80&26.52& 1.23& 0.40& 0.90& 5.50& 1.13\\
           & 100K
&    3&   10&    0& 1.79& 1.85& 1.70&26.61& 1.17& 0.41& 0.91& 5.50& 0.89\\
           & 500K
&   12&   20&   10& 1.64& 1.66& 1.60&26.93& 1.08& 0.58& 1.16& 5.74& 0.81\\
           &   1M
&   24&   30&   20& 1.60& 1.61& 1.58&26.61& 1.06& 0.64& 1.29& 5.83& 0.76\\
           &   2M
&   44&   50&   40& 1.57& 1.58& 1.56&26.96& 1.04& 0.68& 1.41& 5.81& 0.73\\
           &   4M
&   87&   90&   80& 1.55& 1.56& 1.54&26.63& 1.03& 0.69& 1.45& 6.26& 0.72\\
           &   8M
&  167&  171&  160& 1.54& 1.54& 1.53&25.81& 1.02& 0.75& 1.55& 5.98& 0.71\\
           &  16M
&  331&  341&  330& 1.53& 1.53& 1.52&26.75& 1.01& 0.82& 1.70& 6.12& 0.71\\
onezero    &  50K
&    1&   11&    0& 1.50& 1.50& 1.50& 0.01& 1.00& 0.18& 0.14& 1.10& 0.86\\
           & 100K
&    4&   10&    0& 1.50& 1.50& 1.50& 0.02& 1.03& 0.18& 0.15& 1.14& 0.74\\
           & 500K
&   15&   20&   10& 1.50& 1.50& 1.50& 0.00& 1.00& 0.16& 0.15& 1.18& 0.72\\
           &   1M
&   29&   31&   20& 1.50& 1.50& 1.50& 0.01& 1.00& 0.14& 0.14& 1.35& 0.71\\
           &   2M
&   58&   61&   50& 1.50& 1.50& 1.50& 0.01& 1.00& 0.14& 0.14& 1.30& 0.70\\
           &   4M
&  118&  121&  110& 1.50& 1.50& 1.50& 0.01& 1.00& 0.13& 0.13& 1.25& 0.69\\
           &   8M
&  234&  241&  230& 1.50& 1.50& 1.50& 0.01& 1.00& 0.13& 0.13& 1.25& 0.69\\
           &  16M
&  470&  471&  461& 1.50& 1.50& 1.50& 0.02& 1.00& 0.19& 0.18& 1.15& 0.70\\
sorted     &  50K
&    1&   10&    0& 1.89& 2.22& 1.75&26.45& 1.26& 0.41& 0.91& 5.97& 1.15\\
           & 100K
&    2&   10&    0& 1.80& 1.87& 1.64&28.32& 1.18& 0.42& 0.92& 6.16& 0.90\\
           & 500K
&    8&   11&    0& 1.64& 1.66& 1.61&26.84& 1.08& 0.60& 1.20& 6.00& 0.81\\
           &   1M
&   14&   20&   10& 1.60& 1.61& 1.58&26.41& 1.05& 0.66& 1.32& 5.94& 0.76\\
           &   2M
&   26&   30&   20& 1.58& 1.59& 1.57&27.96& 1.04& 0.68& 1.41& 5.89& 0.73\\
           &   4M
&   47&   51&   40& 1.55& 1.56& 1.54&26.72& 1.03& 0.69& 1.45& 6.17& 0.72\\
           &   8M
&   91&  100&   90& 1.54& 1.54& 1.53&25.89& 1.02& 0.73& 1.53& 6.02& 0.71\\
           &  16M
&  179&  190&  170& 1.53& 1.53& 1.52&26.03& 1.01& 0.83& 1.71& 6.19& 0.71\\
organpipe  &  50K
&    0&    0&    0& 1.90& 2.18& 1.81&26.85& 1.24& 0.40& 0.89& 5.17& 1.15\\
           & 100K
&    2&   10&    0& 1.78& 1.88& 1.71&26.20& 1.17& 0.41& 0.90& 5.82& 0.89\\
           & 500K
&    8&   10&    0& 1.64& 1.67& 1.61&27.19& 1.08& 0.58& 1.16& 5.85& 0.81\\
           &   1M
&   16&   20&   10& 1.60& 1.61& 1.59&26.05& 1.06& 0.64& 1.29& 5.88& 0.76\\
           &   2M
&   31&   40&   30& 1.57& 1.58& 1.55&26.99& 1.04& 0.67& 1.40& 6.08& 0.73\\
           &   4M
&   59&   61&   50& 1.55& 1.56& 1.54&25.59& 1.03& 0.69& 1.44& 6.05& 0.72\\
           &   8M
&  116&  121&  110& 1.54& 1.54& 1.53&26.63& 1.02& 0.71& 1.49& 6.23& 0.71\\
           &  16M
&  228&  240&  220& 1.53& 1.53& 1.52&25.67& 1.01& 0.83& 1.71& 5.96& 0.71\\
\hline
\end{tabular}
\end{center}
\end{table}
where the average, maximum and minimum solution times are in
milliseconds, and the comparison counts are in multiples of $n$; e.g.,
column six gives $C_{\rm avg}/n$, where $C_{\rm avg}$ is the
{\em average number of comparisons\/} made over all instances.  Thus
$\gamma_{\rm avg}:=(C_{\rm avg}-1.5n)/f(n)$ estimates the constant
$\gamma$ in the bound \eqref{Cnk}; moreover, for large $n$ we have
$C_{\rm avg}\approx1.5L_{\rm avg}$, where $L_{\rm avg}$ is the average
sum of sizes of partitioned arrays.  Further,
$P_{\rm avg}$ is the average number of {\sc Select} partitions, whereas
$N_{\rm avg}$ is the average number of calls to sSelect and
$p_{\rm avg}$ is the average number of sSelect partitions per call;
both $P_{\rm avg}$ and $N_{\rm avg}$ grow slowly with $\ln n$.
Finally, $s_{\rm avg}$ is the {\em average number of sampled
elements\/}; as predicted by \eqref{sbarletaint},
$s_{\rm avg}/n$ is about $r^{-2}\approx0.69\%$ for large $n$.
The average solution times grow linearly with $n$ (except for small
inputs whose solution times couldn't be measured accurately), and the
differences between maximum and minimum times are quite small (and also
partly due to the operating system).  Except for the smallest inputs,
the maximum and minimum numbers of comparisons are quite close, and
$C_{\rm avg}$ nicely approaches the theoretical lower bound of $1.5n$;
this is reflected in the values of $\gamma_{\rm avg}$ (which are
amazingly stable).  The results for the onezero inputs agree completely
with our theoretical predictions.

For our parameters $\alpha=0.5$ and $\bar\eta=2/r^2$, the test
\eqref{n>=r2} is equivalent to $n\ge r^4$, so \eqref{s1int} operates
only for small $n<r^4=20{,}736$.  Table \ref{tab:Seleta=1/r2}
%
%   *** TABLE 6.2 ***
\begin{table}[t!]
\caption{Performance of {\sc Select} with $\bar\eta=1.000001/r^2$ on
random inputs.}
\label{tab:Seleta=1/r2}
\footnotesize
\begin{center}
\begin{tabular}{lrrrrrrrrrrrrr}
\hline
Input &\multicolumn{1}{c}{Size}
&\multicolumn{3}{c}{Time $[{\rm msec}]$%
\vphantom{$1^{2^3}$}} % Need more vertical space!
&\multicolumn{3}{c}{Comparisons $[n]$}
&\multicolumn{1}{c}{$\gamma_{\rm avg}$}
&\multicolumn{1}{c}{$L_{\rm avg}$}
&\multicolumn{1}{c}{$P_{\rm avg}$}
&\multicolumn{1}{c}{$N_{\rm avg}$}
&\multicolumn{1}{c}{$p_{\rm avg}$}
&\multicolumn{1}{c}{$s_{\rm avg}$}\\
&\multicolumn{1}{c}{$n$}
&\multicolumn{1}{c}{avg}&\multicolumn{1}{c}{max}&\multicolumn{1}{c}{min}
&\multicolumn{1}{c}{avg}&\multicolumn{1}{c}{max}&\multicolumn{1}{c}{min}
& &\multicolumn{1}{c}{$[n]$}
&\multicolumn{1}{c}{$[\ln n]$}
&\multicolumn{1}{c}{$[\ln n]$} &
&\multicolumn{1}{c}{$[\%n]$}\\
\hline
%dsel30a/dsel30ax
random     &  50K
&    7&   11&    0& 2.03& 2.10& 1.94&35.84& 1.35& 0.66& 1.41& 6.43&64.99\\
           & 100K
&   11&   20&   10& 1.82& 1.89& 1.76&29.49& 1.21& 0.65& 1.38& 6.48&45.92\\
           & 500K
&   41&   50&   40& 1.62& 1.64& 1.60&22.69& 1.07& 0.77& 1.62& 6.37&20.48\\
           &   1M
&   70&   91&   60& 1.58& 1.59& 1.56&20.64& 1.05& 0.80& 1.66& 6.37&14.45\\
           &   2M
&  106&  111&  100& 1.55& 1.56& 1.54&18.75& 1.03& 0.87& 1.81& 6.10&10.22\\
           &   4M
&  175&  181&  170& 1.54& 1.54& 1.53&19.07& 1.02& 1.14& 2.34& 6.27& 7.94\\
           &   8M
&  292&  301&  290& 1.53& 1.53& 1.52&18.87& 1.02& 1.32& 2.70& 6.17& 5.81\\
           &  16M
&  498&  501&  491& 1.52& 1.52& 1.52&18.42& 1.01& 1.34& 2.75& 6.40& 4.03\\
\hline
\end{tabular}
\end{center}
\end{table}
highlights the danger of choosing $s_1$ by \eqref{s1int} alone
(note that for $\bar\eta=1.000001r^{-2}$, \eqref{n>=r2} couldn't hold,
being equivalent to $n\ge10^{12}r^4$).  Although $s_{\rm avg}$
increased quite dramatically (cf.\ Tab.\ \ref{tab:Selrand}),
$C_{\rm avg}$ decreased slightly for larger $n$ only,
$\gamma_{\rm avg}$ was less stable and the computing times grew
significantly; similar deteriorations occured for other inputs.

Although it is not clear how to implement the theoretical scheme of
Knuth \cite[Ex.\ 5.3.3--24]{knu:acpIII2}, we tried to emulate it by
using $r^2=2$ and \eqref{glsmall} replaced for $l\le\bar l$ by
\begin{equation}
g_l:=\left(\,\min\{\theta,1-\theta\} s_l\ln s_l\,\right)^{1/2}.
\label{glK}
\end{equation}
Relative to Tab.\ \ref{tab:Selrand}, this scheme made about $3\%$ more
comparisons for small $n$, but was about 9.5 times slower due to the
random sampling overheads (with $s_{\rm avg}$ between $52\%$ and $57\%$).
Eliminating randomization
% (i.e., the swaps of Steps 2 and 3; cf.\ \S\ref{s:impl})
gave the results of Table \ref{tab:SelK}.
%
%   *** TABLE 6.3 ***
\begin{table}[h]
\caption{Performance of {\sc Select} with Knuth's gap \eqref{glK}
and no randomization.}
\label{tab:SelK}
\footnotesize
\begin{center}
\tabcolsep=0.90\tabcolsep
\begin{tabular}{lrrrrrrrrrrrr}
\hline
Input &\multicolumn{1}{c}{Size}
&\multicolumn{3}{c}{Time $[{\rm msec}]$%
\vphantom{$1^{2^3}$}} % Need more vertical space!
&\multicolumn{3}{c}{Comparisons $[n]$}
&\multicolumn{1}{c}{$\gamma_{\rm avg}$}
&\multicolumn{1}{c}{$L_{\rm avg}$}
&\multicolumn{1}{c}{$P_{\rm avg}$}
&\multicolumn{1}{c}{$N_{\rm avg}$}
&\multicolumn{1}{c}{$p_{\rm avg}$}\\
&\multicolumn{1}{c}{$n$}
&\multicolumn{1}{c}{avg}&\multicolumn{1}{c}{max}&\multicolumn{1}{c}{min}
&\multicolumn{1}{c}{avg}&\multicolumn{1}{c}{max}&\multicolumn{1}{c}{min}
& &\multicolumn{1}{c}{$[n]$}
&\multicolumn{1}{c}{$[\ln n]$}
&\multicolumn{1}{c}{$[\ln n]$} & \\
\hline
%dsel31a/dsel31ax
random     &  50K
&    4&   10&    0& 1.99& 2.15& 1.87&32.98& 1.42& 3.35& 6.08& 5.18\\
           & 100K
&    4&   10&    0& 1.86& 2.09& 1.77&33.13& 1.31& 4.40& 7.95& 4.95\\
           & 500K
&   15&   20&   10& 1.67& 2.01& 1.63&32.55& 1.14& 7.09&12.65& 5.01\\
           &   1M
&   33&   41&   30& 1.67& 2.01& 1.59&44.80& 1.15& 8.84&15.49& 5.03\\
           &   2M
&   60&   70&   50& 1.61& 1.81& 1.56&39.10& 1.09& 9.23&16.57& 5.29\\
           &   4M
&  118&  121&  110& 1.57& 1.67& 1.55&33.66& 1.06&12.51&21.86& 5.08\\
           &   8M
&  244&  300&  240& 1.55& 1.81& 1.53&34.39& 1.04&13.95&24.56& 5.16\\
           &  16M
&  493&  601&  460& 1.58& 1.81& 1.52&81.48& 1.08&18.07&30.75& 5.09\\
onezero    &   8M
&  297&  301&  290& 1.50& 1.50& 1.50& 0.09& 1.00& 1.45& 0.19& 1.15\\
           &  16M
&  582&  591&  580& 1.50& 1.50& 1.50& 0.11& 1.00& 1.45& 0.18& 1.10\\
%twofaced   &   8M
%&  247&  291&  240& 1.56& 1.78& 1.53&43.50& 1.05&14.82&25.35& 5.04\\
%           &  16M
%&  478&  591&  461& 1.55& 1.77& 1.52&50.32& 1.05&18.03&30.93& 5.07\\
sorted     &  50K
&   23&   30&   20&46.19&46.19&46.19&  ***&39.86&216.3&366.0& 5.18\\
           & 100K
&   56&   61&   50&56.16&56.16&56.16&  ***&48.59&471.0&776.0& 5.16\\
           & 500K
&  410&  421&  400&85.83&85.83&85.83&  ***&75.16&  ***&  ***& 5.37\\
%           &   1M
%&  976&  991&  971&  ***&  ***&  ***&  ***&89.41&  ***&  ***& 5.26\\
%           &   2M
%& 2356& 2384& 2343&  ***&  ***&  ***&  ***&106.8&  ***&  ***& 5.28\\
%           &   4M
%& 5605& 5648& 5568&  ***&  ***&  ***&  ***&123.8&  ***&  ***& 5.28\\
           &   8M
&13625&13690&13579&  ***&  ***&  ***&  ***&147.7&  ***&  ***& 5.29\\
           &  16M
&32095&32186&31986&  ***&  ***&  ***&  ***&175.7&  ***&  ***& 5.42\\
%rotated    &  50K
%&   23&   30&   20&45.73&45.73&45.73&  ***&39.41&215.8&363.0& 5.14\\
%           & 100K
%&   55&   61&   50&55.85&55.85&55.85&  ***&48.28&472.8&780.0& 5.16\\
%           & 500K
%&  409&  421&  400&86.23&86.23&86.23&  ***&75.53&  ***&  ***& 5.37\\
%           &   1M
%&  974&  982&  961&  ***&  ***&  ***&  ***&89.94&  ***&  ***& 5.30\\
organpipe  &   8M
& 7238& 7281& 7200&81.08&81.08&81.08&  ***&71.59&  ***&  ***& 5.06\\
           &  16M
&16486&16564&16453&90.76&90.76&90.76&  ***&80.55&  ***&  ***& 5.18\\
%m3killer   &   8M
%& 1660& 1703& 1632&15.80&15.80&15.80&  ***&12.66&  ***&  ***& 5.22\\
%           &  16M
%& 3968& 4006& 3936&17.48&17.48&17.48&  ***&14.29&  ***&  ***& 5.56\\
\hline
\end{tabular}
\end{center}
\end{table}
Not suprisingly, this scheme performed fairly well on the random inputs,
% (with slowdowns of up to $48\%$ relative to Tab.\ \ref{tab:Selrand}),
but quite badly on the deterministic inputs (where ``***'' denote
values exceeding the printout format).

Finally, comparing Tab.\ \ref{tab:Selrand} with
\cite[Tabs.\ 7.1--7.2]{kiw:rsq}, we add that {\sc Select} was slightly
slower than its counterpart of \cite{kiw:rsq},
% by about $5\%$ to $18\%$,
although the numbers of comparisons made were similar for large $n$.
In fact for small inputs, the ternary version of \cite{kiw:rst} made
fewest comparisons.  The experimental results of \cite{kiw:psq,kiw:rsq}
suggest that {\sc Select} can compete successfully with refined
implementations of quickselect.

%{\bf Acknowledgment}.  I would like to thank the Associate Editor and
%the two anonymous referees for their helpful comments.
{\bf Acknowledgment}.  I would like to thank Olgierd Hryniewicz,
Roger Koenker, Ronald L. Rivest and John D. Valois for useful
discussions.

%
%   *** REFERENCES ***
\footnotesize
%\bibliography{kckabbr,kalg,kbk,kck,kint,kth}
%\bibliographystyle{kck}
\newcommand{\etalchar}[1]{$^{#1}$}
\newcommand{\noopsort}[1]{} \newcommand{\printfirst}[2]{#1}
  \newcommand{\singleletter}[1]{#1} \newcommand{\switchargs}[2]{#2#1}
\ifx\undefined\bysame
\newcommand{\bysame}{\leavevmode\hbox to3em{\hrulefill}\,}
\fi

\normalsize
%   *** END OF REFERENCES ***
%
\end{document}